\documentclass{PoS}

\usepackage{cite}
\usepackage{xspace}
\usepackage{amsmath}
\RequirePackage{booktabs}
\RequirePackage{subcaption}
\RequirePackage{multirow}

\def\asymp#1%
{\mathrel{\raisebox{-.4em}{$\widetilde{\scriptstyle #1}$}}}

\def\Nequal#1%
{\mathrel{\raisebox{-.5em}{$\stackrel{=}{\scriptstyle\rm#1}$}}}
\newcommand{\dsl}[1]{\not \hspace{-0.7mm}#1}
\def\dsl{\mathpalette\make@slash}
\def\make@slash#1#2{\setbox\z@\hbox{$#1#2$}%
  \hbox to 0pt{\hss$#1/$\hss\kern-\wd0}\box0}

\def\beq{\begin{equation}}
\def\eeq{\end{equation}}
\def\bit{\begin{itemize}}
\def\eit{\end{itemize}}
\def\beqar{\begin{eqnarray}}
\def\eeqar{\end{eqnarray}}
\def\barr#1{\begin{array}{#1}}
\def\earr{\end{array}}
\def\bfi{\begin{figure}}
\def\efi{\end{figure}}
\def\btab{\begin{table}}
\def\etab{\end{table}}
\def\bce{\begin{center}}
\def\ece{\end{center}}

\newcommand{\GeV}{\unskip\,\mathrm{GeV}}
\newcommand{\MeV}{\unskip\,\mathrm{MeV}}

\newcommand{\ri}{{\mathrm{i}}}

\newcommand{\rT}{{\mathrm{T}}}

\def\mathswitchr#1{\relax\ifmmode{\mathrm{#1}}\else$\mathrm{#1}$\fi}

\newcommand{\PV}{V}
\newcommand{\PW}{\mathswitchr W}

\newcommand{\PZ}{\mathswitchr Z}

\newcommand{\Pp}{\mathswitchr p}

\newcommand{\PWp}{\mathswitchr {W^+}}

\newcommand{\jet}{\mathswitchr {jet}}

\newcommand{\Pl}{\ell}

\def\mathswitch#1{\relax\ifmmode#1\else$#1$\fi}

\newcommand{\MV}{\mathswitch {M_\PV}}
\newcommand{\MW}{\mathswitch {M_\PW}}

\newcommand{\MZ}{\mathswitch {M_\PZ}}

\newcommand{\GW}{\Gamma_{\PW}}

\def\solid{\raise.9mm\hbox{\protect\rule{1.1cm}{.2mm}}}
\def\dash{\raise.9mm\hbox{\protect\rule{2mm}{.2mm}}\hspace*{1mm}}

\newcommand{\LO}{{\mathrm{LO}}}
\newcommand{\NLO}{{\mathrm{NLO}}}

\newcommand{\alphas}{\alpha_\mathrm{s}}

\newcommand{\order}[1]{\ensuremath{{\mathcal{O}\!\left(#1\right)}}\xspace}

\DeclareRobustCommand{\ensuremathrm}[1]{\ensuremath{\mathrm{#1}}}

\DeclareRobustCommand{\pro}{\text{prod}\xspace}
\DeclareRobustCommand{\dec}{\text{dec}\xspace}
\DeclareRobustCommand{\LO}{\text{LO}\xspace}
\DeclareRobustCommand{\NLO}{\text{NLO}\xspace}
\DeclareRobustCommand{\NNLO}{\text{NNLO}\xspace}

\DeclareRobustCommand{\rs}{\ensuremathrm{s}}
\DeclareRobustCommand{\rew}{\ensuremathrm{ew}}

\DeclareRobustCommand{\SI}[2]{\ensuremath{{#1}~{#2}}\xspace}

\DeclareRobustCommand{\OS}{\text{OS}\xspace}

\title{Dominant {\boldmath$\order{\alpha_{\mathrm{s}}\alpha}$}
corrections to Drell--Yan processes in the resonance region}

\ShortTitle{$\order{\alpha_{\mathrm{s}}\alpha}$ corrections to Drell--Yan processes}

\author{{Stefan Dittmaier}\\
         Albert-Ludwigs-Universit\"at Freiburg, Physikalisches Institut,
        D-79104 Freiburg, Germany\\
        E-mail: \email{stefan.dittmaier@physik.uni-freiburg.de}}

\author{{Alexander Huss}\\
        Institute for Theoretical Physics, ETH, CH-8093 Z\"urich, Switzerland\\
        Department of Physics, University of Z\"urich, CH-8057 Z\"urich, Switzerland\\
        E-mail: \email{ahuss@phys.ethz.ch}}

\author{Christian Schwinn\\
        Institute for Theoretical Particle Physics and Cosmology,
        RWTH Aachen University, D-52056 Aachen, Germany\\
        E-mail: \email{schwinn@physik.rwth-aachen.de}}

\abstract{%
Apart from the well-known NNLO QCD and NLO electroweak corrections to
W- and Z-boson production at hadron colliders, the most important
fixed-order corrections
are given by the mixed QCD--electroweak corrections of
$\order{\alphas\alpha}$.
The knowledge of these corrections is of particular importance
to control the theoretical uncertainties
in the upcoming high-precision measurements of the W-boson mass
and the effective weak mixing angle at the LHC.
Since these observables are dominated by the phase-space regions
of resonant W/Z~bosons, we address the $\order{\alphas\alpha}$
corrections in the framework of an expansion about the
W/Z~poles.
Retaining only the leading, resonant contribution in the so-called pole approximation,
the corrections can be classified into factorizable and non-factorizable contributions.
In this article we review our calculation of the
numerically dominant corrections which arise from
factorizable corrections of ``initial--final'' type, i.e.\
they combine the QCD corrections to the production with the large
electroweak corrections to the decay of the W/Z~boson.
Moreover, we compare our results to simpler approximate combinations of
electroweak and QCD corrections based on
naive products of NLO QCD and
electroweak correction factors and using  leading-logarithmic approximations
for QED final-state radiation.
Finally, we estimate the shift in the W-boson mass that results from the
$\order{\alphas\alpha}$ corrections to the transverse-mass distribution.
}

\FullConference{12th International Symposium on Radiative Corrections (Radcor 2015) and LoopFest XIV (Radiative Corrections for the LHC and Future Colliders)\\
		15-19 June, 2015\\
		UCLA Department of Physics \& Astronomy Los Angeles, USA}

\begin{document}

\section{Introduction}

The Drell--Yan-like production of W and Z~bosons,
$\Pp\Pp/\Pp\bar\Pp \to V \to \Pl_1\bar\Pl_2 + X$,
is one of the most prominent classes of particle reactions at hadron colliders.
The large production rate and the clean experimental signature
of the leptonic vector-boson decay allow these processes to be measured with great precision
and render them one of the most important ``standard-candle'' processes at the LHC.
Not only do these processes represent powerful tools for detector calibration, but they can also be used as luminosity monitor and further deliver important constraints in the fit of parton distribution functions (PDFs).
Of particular relevance for precision tests of the Standard Model
is the potential of the Drell--Yan process at the LHC for high-precision measurements
in the resonance regions.
In particular, the effective weak mixing angle might be measured with LEP precision and the W-boson mass is expected to be extracted from kinematic fits with a sensitivity below \SI{10}{\MeV} (see Ref.~\cite{Baak:2013fwa} and references therein).

On the theory side, the Drell--Yan-like production of W or Z bosons is one of the best understood and most precisely predicted processes.
The current state of the art includes
QCD corrections at next-to-next-to-leading-order (NNLO) accuracy,
supplemented by leading higher-order soft-gluon effects
or matched to QCD parton showers up to NNLO.
The electroweak (EW) corrections are known at next-to-leading order (NLO)
and leading universal corrections beyond
(see, e.g., references in Ref.~\cite{Dittmaier:2015rxo}).
Thus, in addition to the N$^3$LO QCD corrections, the next frontier in theoretical
fixed-order computations is given by the calculation of the mixed QCD--EW
corrections of \order{\alphas\alpha},
which can affect observables relevant for the W-mass~determination at the percent level.

While leading universal QCD and EW corrections are known to factorize from each other,
a full NNLO calculation at \order{\alphas\alpha} is necessary
for a proper combination of NLO QCD and NLO EW corrections without ambiguities.
Here some partial results for two-loop
amplitudes~\cite{Kotikov:2007vr,Kilgore:2011pa,Bonciani:2011zz} as well as the full
\order{\alphas\alpha} corrections to the W/Z decay widths~\cite{Czarnecki:1996ei,Kara:2013dua}
are known.
A complete calculation of the \order{\alphas\alpha} corrections requires to combine
the double-virtual corrections with the \order{\alpha} EW corrections to $\PW/\PZ+\jet$
production, the \order{\alphas} QCD corrections to $\PW/\PZ+\gamma$
production, and the double-real corrections.

In a series of two recent papers~\cite{Dittmaier:2014qza,Dittmaier:2015rxo},
we have initiated the calculation of the \order{\alphas\alpha} corrections
to Drell--Yan processes in the resonance region via the so-called
\emph{pole approximation}~(PA).
It is based on a systematic expansion of the cross section about the
resonance pole and is suitable for theoretical predictions in the
vicinity of the gauge-boson resonance.
In detail, the PA splits the corrections into factorizable and non-factorizable contributions. The former
can be separately attributed to the production and
the subsequent decay of the gauge boson, while the latter link
the production and decay subprocesses by the exchange of soft photons.

In this article we motivate the general idea of the pole expansion and outline the salient features of the PA at $\order{\alphas\alpha}$.
We discuss our numerical results for the factorizable corrections of ``initial--final'' type, which are the dominant contribution at this order, as they combine sizable QCD corrections to the production with the large EW corrections to the W/Z~decays.
Finally, we present the impact of the $\order{\alphas\alpha}$ corrections on the W-boson mass extraction from a kinematic fit to the transverse-mass spectrum.

\section{Structure of the pole approximation}

The PA for Drell--Yan processes
provides a systematic classification of contributions to Feynman
diagrams that are enhanced by the resonant propagator of a vector
boson~$V=\PW,\PZ$.
To this end, we schematically write the transition amplitude of the process in the following form,
\begin{equation}
  \mathcal{M} = \frac{W(p_V^2)}{p_V^2-M_V^2+\Sigma(p_V^2)} + N(p_V^2) ,
\end{equation}
where $\Sigma$ denotes the self-energy of $V$ and the functions $W$ and $N$ represent the resonant and non-resonant parts, respectively.
In order to isolate the resonant contributions in a gauge-invariant way, we further rewrite the above expression as
\begin{equation}
  \label{eq:PA}
  \mathcal{M} = \frac{W(\mu_V^2)}{p_V^2-\mu_V^2}\;\frac{1}{1+\Sigma'(\mu_V^2)}
  + \left[ \frac{W(p_V^2)}{p_V^2-M_V^2+\Sigma(p_V^2)}
         - \frac{W(\mu_V^2)}{p_V^2-\mu_V^2}\;\frac{1}{1+\Sigma'(\mu_V^2)} \right]
  + N(p_V^2)
\end{equation}
with $\mu_V^2=M_V^2-\ri M_V\Gamma_V$ denoting the gauge-invariant location of the propagator pole in the complex $p_V^2$ plane.

The amplitude in the PA is obtained from Eq.~\eqref{eq:PA} by omitting the last, non-resonant term and systematically expanding the term in square brackets about the point $p_V^2=M_V^2$ and only keeping the leading, resonant contribution.
The first term in Eq.~\eqref{eq:PA} corresponds to the \emph{factorizable} corrections in which on-shell production and decay subamplitudes are linked by the off-shell propagator.
The evaluation of the subamplitudes using on-shell kinematics is essential in order to guarantee gauge invariance.
The specification of the on-shell projection is not unique and different variants lead to differences within the intrinsic accuracy of the PA which is of $\order{\tfrac{\alpha}{\pi}\times\tfrac{\Gamma_V}{M_V}}$ in cross-section contributions that correct the LO prediction by terms of $\order{\alpha}$.
Despite the freedom in the choice of the on-shell projection, they have to match between the virtual and real corrections in the infrared-singular limits to ensure the proper cancellation of the singularities in the final result.
The \emph{non-factorizable} corrections arise from the term on the r.h.s.\ of Eq.~\eqref{eq:PA} in square brackets and are deeply linked to the soft-singular structure of $W(p_V^2)$ and $\Sigma'(p_V^2)$ in the limit $p_V^2\to\mu_V^2$.
As explained in more detail in Ref.~\cite{Dittmaier:2014qza}, these corrections solely arise from soft-photon exchange that link production and decay.

As a result, the corrections to the production and decay stages of the intermediate unstable particle are separated in a consistent and gauge-invariant way by the PA.
This is
particularly relevant for the charged-current Drell-Yan process, where
photon radiation off the intermediate $\PW$~boson contributes
simultaneously to the corrections to production and decay of a
$\PW$~boson, and to the non-factorizable contributions.  Applications
of different variants of the PA to NLO EW
corrections~\cite{Baur:1998kt,Dittmaier:2001ay,Dittmaier:2014qza}
have been validated by a comparison to the complete NLO EW
calculations and show excellent agreement at the order of some $0.1\%$
in kinematic distributions dominated by the resonance region.
In particular, the bulk of the NLO EW corrections near the resonance
can be attributed to the factorizable corrections to the W/Z~decay subprocesses,
while the factorizable corrections to the production process are
mostly suppressed below the percent level,
and the non-factorizable contributions are even smaller.

\begin{figure}
  \centering
  \begin{subfigure}[m]{.48\linewidth}
    \centering
    \includegraphics{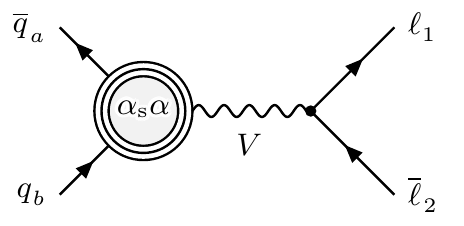}
    \subcaption{Factorizable initial--initial corrections}
  \end{subfigure}
  \begin{subfigure}[m]{.48\linewidth}
    \centering
    \includegraphics{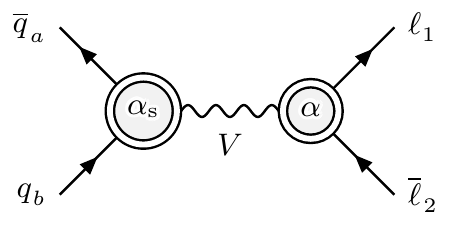}
    \subcaption{Factorizable initial--final corrections}
  \end{subfigure}
  \\[1.5em]
  \begin{subfigure}[m]{.48\linewidth}
    \centering
    \includegraphics{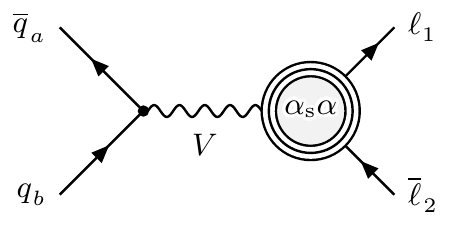}
    \subcaption{Factorizable final--final corrections}
  \end{subfigure}
  \begin{subfigure}[m]{.48\linewidth}
    \centering
    \includegraphics{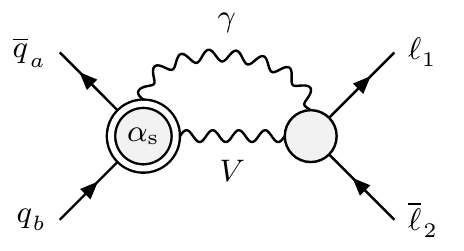}
    \subcaption {Non-factorizable corrections}
  \end{subfigure}
  \caption{The four types of corrections that contribute to the mixed QCD--EW corrections in the PA illustrated in terms of generic two-loop amplitudes. Simple circles symbolize tree structures, double circles one-loop corrections, and triple circles two-loop contributions.}
  \label{fig:NNLOcontrib}
\end{figure}

The quality of the PA at NLO justifies the application of this approach
in the calculation of the $\order{\alphas\alpha}$ corrections to
observables that are dominated by the resonances.
The structure of the PA for
the \order{\alphas\alpha} correction has
been worked out in Ref.~\cite{Dittmaier:2014qza}, where details of the
method and our setup can be found.
The corrections can be classified into the four types of
contributions shown in Fig.~\ref{fig:NNLOcontrib} for the case of the
double-virtual corrections. For each class of contributions with
the exception of the final--final corrections (c), also the
associated real--virtual and
double-real corrections have to be
computed, obtained by replacing one or both of the labels $\alpha$ and $\alphas$ in
the blobs in Fig.~\ref{fig:NNLOcontrib} by a real
photon or gluon, respectively. The corresponding crossed partonic channels, e.g.\ with
quark--gluon initial states have to be included in addition.

In detail, the four types of corrections are characterized as follows:
\begin{enumerate}
\renewcommand{\labelenumi}{(\alph{enumi})}
\item The initial--initial factorizable corrections are given by
  two-loop \order{\alphas\alpha} corrections to on-shell $\PW/\PZ$
  production and the corresponding one-loop real--virtual and
  tree-level double-real contributions, i.e.\ $\PW/\PZ+\jet$
  production at \order{\alpha}, $\PW/\PZ+\gamma$ production at
  \order{\alphas}, and the processes $\PW/\PZ+\gamma+\jet$ at tree
  level.  Results for individual ingredients of the initial--initial
  part are known, however, a consistent combination of these building blocks requires
  also a subtraction scheme for infrared (IR) singularities at $\order{\alphas\alpha}$
  and has not been performed yet.
  Note that
  currently no PDF set including \order{\alphas\alpha} corrections is
  available, which is required to absorb IR singularities of the
  initial--initial corrections from QCD and photon radiation collinear to the
  beams.

  Results of the PA at \order{\alpha} show that observables such as
  the transverse-mass distribution in the case of W~production or
  the lepton-invariant-mass distributions for Z~production are
  extremely insensitive to photonic initial-state
  radiation (ISR)~\cite{Dittmaier:2014qza}. Since these distributions also
  receive relatively moderate QCD corrections, we do not expect
  significant initial--initial NNLO
  \order{\alphas\alpha}
  corrections to such distributions. For observables sensitive to
  initial-state recoil effects, such as the transverse-lepton-momentum
  distribution,
  the \order{\alphas\alpha} corrections should be larger, but still
  very small compared to the huge QCD corrections.

\item  The factorizable initial--final corrections
consist of the \order{\alphas} corrections to $\PW/\PZ$ production combined with the \order{\alpha} corrections to the leptonic $\PW/\PZ$ decay.
The latter comprise the by far dominant contribution to the PA at $\order{\alpha}$ with a substantial impact on the shape of differential distributions.
Further given that the NLO QCD correction are sizable with no apparent suppression, we expect the factorizable corrections of ``initial--final'' type to capture the dominant \order{\alphas\alpha} effects.
These NNLO corrections receive double-virtual, real--virtual, and double-real contributions, as illustrated in Fig.~\ref{fig:NNLO-if-graphs} in terms of generic interference diagrams.
The cancellation of IR singularities between these individual contributions is achieved using a local subtraction scheme resulting in a fully differential calculation.
In order to construct our NNLO subtraction terms we employ a two-fold application of the dipole subtraction formalism for NLO QCD and EW corrections.
An important ingredient for this construction was the recent generalization of the formalism to cover decay kinematics, which has been worked out in Ref.~\cite{Basso:2015gca}.
Further details on the computation can be found in Ref.~\cite{Dittmaier:2015rxo}.
In the following sections we review the main results presented there.
\begin{figure}
  \centering
  \begin{subfigure}[m]{\linewidth}
    \centering
    \includegraphics[scale=0.9]{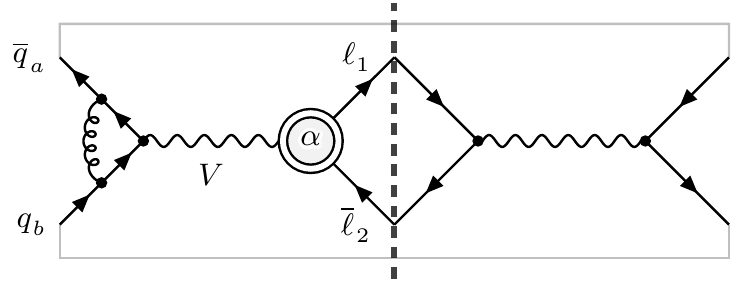}
    \qquad
    \includegraphics[scale=0.9]{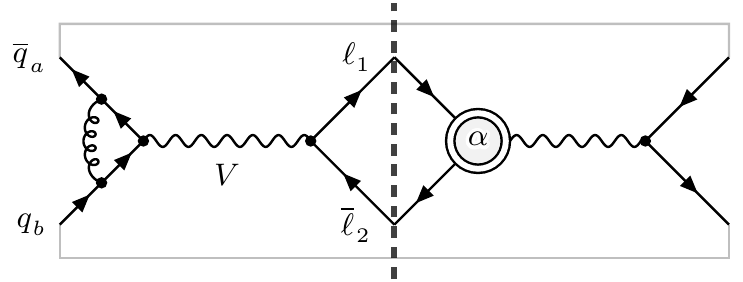}
    \caption{Factorizable initial--final double-virtual corrections}
    \label{fig:NNLO-if-graphs-vv}
  \end{subfigure}
  \\[1.5em]
  \begin{subfigure}[m]{\linewidth}
    \centering
    \includegraphics[scale=0.9]{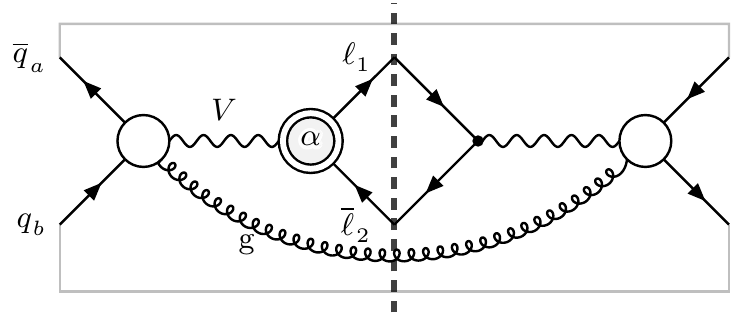}
    \qquad
    \includegraphics[scale=0.9]{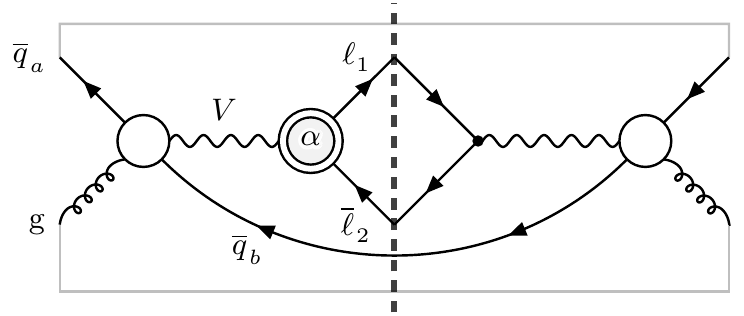}
    \caption{Factorizable initial--final (real QCD)$\times$(virtual EW) corrections}
    \label{fig:NNLO-if-graphs-rv}
  \end{subfigure}
  \\[1.5em]
  \begin{subfigure}[m]{\linewidth}
    \centering
    \includegraphics[scale=0.9]{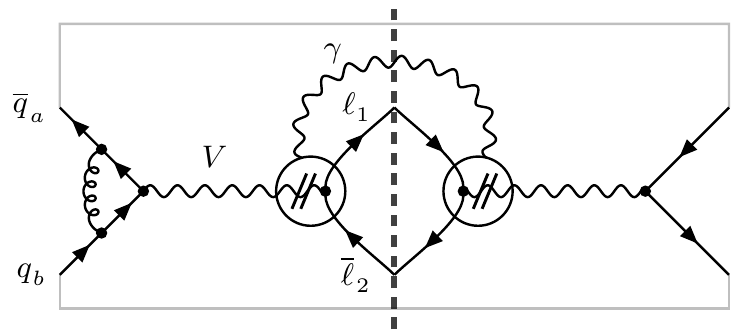}
    \caption{Factorizable initial--final (virtual QCD)$\times$(real photonic) corrections}
    \label{fig:NNLO-if-graphs-vr}
  \end{subfigure}
  \\[1.5em]
  \begin{subfigure}[m]{\linewidth}
    \centering
    \includegraphics[scale=0.9]{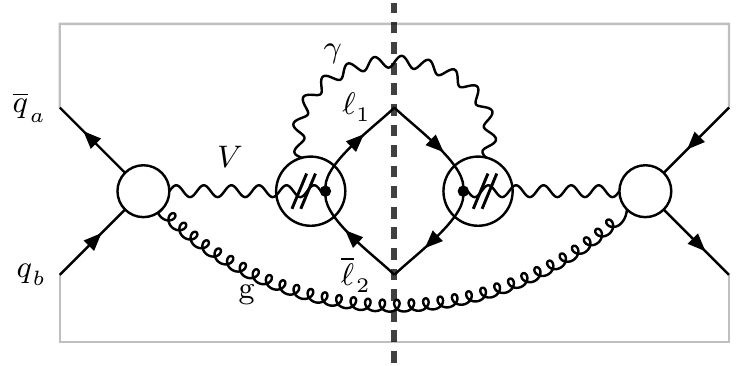}
    \qquad
    \includegraphics[scale=0.9]{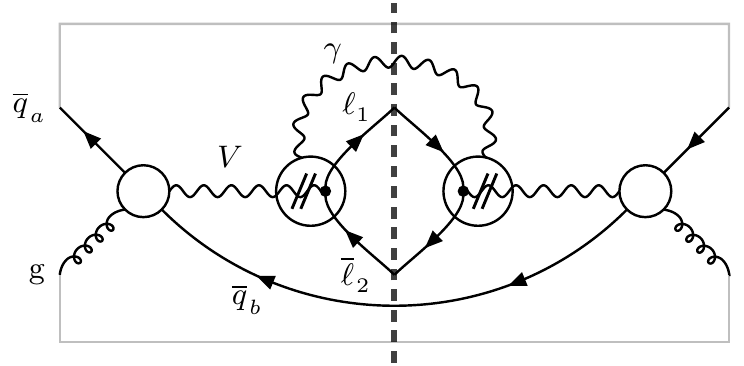}
    \caption{Factorizable initial--final double-real corrections}
    \label{fig:NNLO-if-graphs-rr}
  \end{subfigure}
  \caption{Interference diagrams for the various contributions to the factorizable initial--final corrections of $\order{\alphas\alpha}$, with blobs representing all relevant tree structures.
The blobs with ``$\alpha$'' inside represent one-loop corrections of $\order{\alpha}$,
and the double slash on a propagator line indicates that the corresponding momentum
is set on its mass shell in the rest of the diagram (but not on the slashed line itself).}
\label{fig:NNLO-if-graphs}
\end{figure}

\item Factorizable final--final corrections arise solely from the
  $\order{\alphas\alpha}$ counterterms of the
  lepton--$\PW/\PZ$-boson vertices, which involve only corrections to
  the vector-boson self-energies at this order.
  Not only are these corrections free of IR divergences, but owing to the fact that there are no corresponding real contributions, the final--final corrections have practically no impact on the shape of distributions.
The explicit calculation carried out in Ref.~\cite{Dittmaier:2015rxo} further
reveals that those corrections are in fact phenomenologically negligible.

\item  The non-factorizable $\order{\alphas\alpha}$ corrections are given by soft-photon corrections
connecting the initial state, the intermediate vector boson, and the
final-state leptons, combined with QCD corrections to $V$-boson production.
As shown in detail in Ref.~\cite{Dittmaier:2014qza}, these corrections
can be expressed in terms of soft-photon
correction factors to squared tree-level or one-loop QCD matrix
elements by using gauge-invariance arguments.
Furthermore, exploiting this factorization property, the IR cancellation can be accomplished by using a composition of two NLO methods:
We employ the dipole-subtraction formalism for the treatment of the IR singularities associated with the QCD corrections together with the soft-slicing approach for the photonic corrections.
The numerical impact of these corrections was found to be below the
$0.1\%$ level and is therefore negligible for all phenomenological purposes.
\end{enumerate}

\section{Numerical results}

\subsection{Setup and conventions}

The detailed setup and the input parameters used to obtain the results
discussed in the following can be found in Ref.~\cite{Dittmaier:2015rxo}.
Here we just repeat the basic definition of the individual components
of the QCD and EW corrections.

Our default prediction for the Drell--Yan cross section at mixed QCD--EW NNLO
is obtained by
adding the \order{\alphas\alpha} corrections $\Delta\sigma_{\pro\times\dec}^{\NNLO_{\rs\otimes\rew}}$
to the sum $\Delta\sigma^{\NLO_\rs} + \Delta\sigma^{\NLO_\rew}$ of the full NLO QCD and EW corrections,
\begin{align}
\label{eq:def:nnlo:if}
  \sigma^{{\NNLO}_{\rs\otimes\rew}} &=
  \sigma^0 +\Delta\sigma^{\NLO_\rs} + \Delta\sigma^{\NLO_\rew}
  +\Delta\sigma_{\pro\times\dec}^{\NNLO_{\rs\otimes\rew}} ,
\end{align}
where all terms are consistently evaluated with NLO PDFs including the leading-order (LO) contribution $\sigma^0$.
The non-factorizable corrections as well as the factorizable corrections of ``final--final'' type
are not taken into account due to their negligible size, as discussed above.

In order to validate estimates of the NNLO QCD--EW corrections based on a naive product ansatz,
we define the naive product of the NLO QCD cross section and the relative EW corrections,
\begin{align}
  \sigma_{\text{naive fact}}^{\NNLO_{\rs\otimes\rew}} &=
  \sigma^{\NLO_\rs} (1+\delta_{\alpha})
=
  \sigma^0 +\Delta\sigma^{\NLO_\rs} + \Delta\sigma^{\NLO_\rew}
  +\Delta\sigma^{\NLO_\rs}\;\delta_{\alpha}  .
  \label{eq:def:nnlo:naive:fact}
\end{align}
The relative NLO EW corrections
\begin{align}
  \label{eq:def:deltaew}
  \delta_{\alpha} &\equiv
  \frac{\Delta\sigma^{\NLO_\rew}}{\sigma^0}
\end{align}
are defined in two different versions: First, based on the full \order{\alpha} correction
($\delta_\alpha$), and second, based on the dominant EW final-state correction of the PA
($\delta_\alpha^{\dec}$).

Defining the correction factors,%
\footnote{
  Note that the correction factor $\delta_{\alphas}'$ differs from that in the standard QCD $K$~factor $K_{\mathrm{NLO}_{\mathrm{s}}}=\sigma_{\mathrm{NLO}_{\mathrm{s}}}/\sigma_{\mathrm{LO}}\equiv 1+\delta_{\alpha_s}$ due to the use of different PDF sets in the Born contributions appearing in the normalization.
}
\begin{align}
  \label{eq:def:deltas'}
  \delta^{{\pro}\times{\dec}}_{\alphas\alpha} &\equiv
  \frac{\Delta\sigma_{{\pro}\times{\dec}}^{\NNLO_{{\rs}\otimes{\rew}}}}{\sigma^{\LO}} , &
  \delta_{\alphas}' &\equiv
  \frac{\Delta\sigma^{\NLO_{\rs}}}{\sigma^{\LO}} ,
\end{align}
we can cast the relative difference of our best prediction~\eqref{eq:def:nnlo:if} and the
product ansatz~\eqref{eq:def:nnlo:naive:fact} into the following form,
\begin{align}
\label{eq:diff-naive}
  \frac{\sigma^{\NNLO_{\rs\otimes\rew}}
    -\sigma_{\text{naive fact}}^{\NNLO_{\rs\otimes\rew}}}
  {\sigma^{\LO}}
  &=
  \delta^{{\pro}\times{\dec}}_{\alphas\alpha} - \delta_{\alphas}' \delta_\alpha ,
\end{align}
where the LO prediction $\sigma^{\LO}$ in the denominators is evaluated with the LO PDFs.
The difference of the relative NNLO
correction $\delta^{\pro\times\dec}_{\alphas\alpha}$ and the naive
product $\delta_{\alphas}' \delta_\alpha^{(\dec)}$ therefore allows to
assess the validity of a naive product ansatz.

Most contributions to the
factorizable initial--final corrections take the reducible form of a
product of two NLO corrections, with the exception of the double-real
emission corrections which are defined with the full kinematics of the
$2\to 4$ phase space.
It is only in the double-real contributions where the final-state leptons
receive recoils from both QCD and photonic radiation, an effect that cannot be
captured by naively multiplying NLO QCD and EW corrections.
Any large deviations between
$\delta^{\pro\times\dec}_{\alphas\alpha}$ and
$\delta_{\alphas}'\delta_\alpha^{(\dec)}$ can therefore be attributed
to this type of contribution. The difference of the naive products
defined in terms of $\delta_\alpha^{\dec}$ and $\delta_\alpha$
allows us to assess the  impact of the missing $\order{\alphas\alpha}$ corrections beyond the initial--final
corrections considered in our calculation and therefore also provides
an error estimate of the PA, and in particular of the omission of the
corrections of ``initial--initial'' type.

\subsection{Results on the dominating \order{\alphas\alpha} corrections}

\begin{figure}
  \includegraphics[width=.5\textwidth]{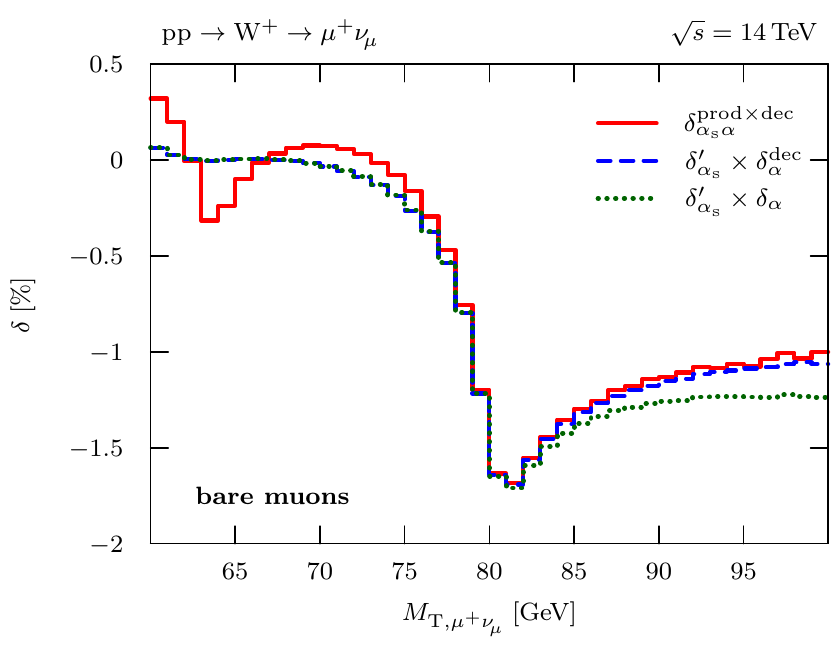}
  \includegraphics[width=.5\textwidth]{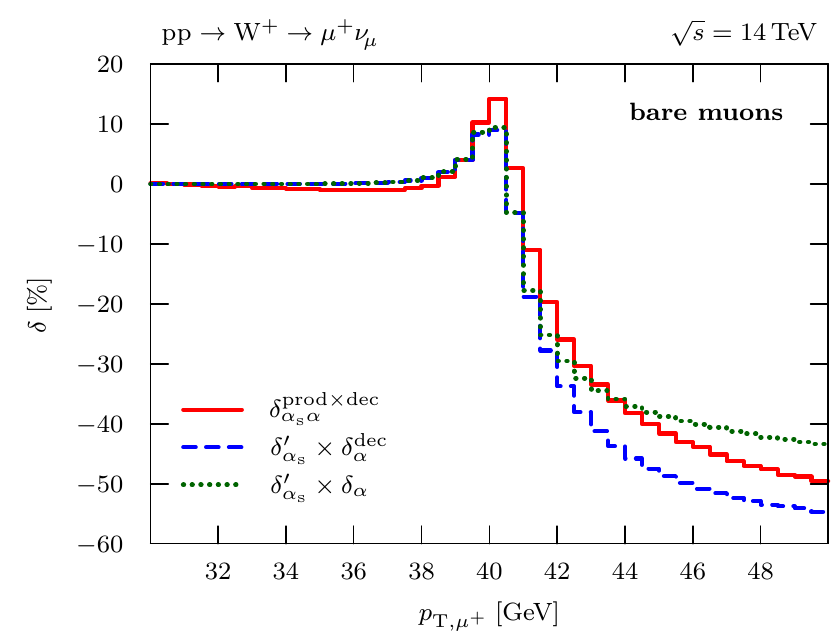}
  \caption{Relative factorizable corrections of $\order{\alphas\alpha}$ induced by initial-state QCD and final-state EW contributions to the transverse-mass~(left) and transverse-lepton-momentum~(right) distributions for \PWp~production at the LHC.
  The naive products of the NLO correction factors $\delta_{\alphas}'$ and $\delta_\alpha$ are shown for comparison. (Taken from Ref.~\cite{Dittmaier:2015rxo}.)}
  \label{fig:distNNLO-IF-Wp}
%
\vspace{2em}
  \includegraphics[width=.5\textwidth]{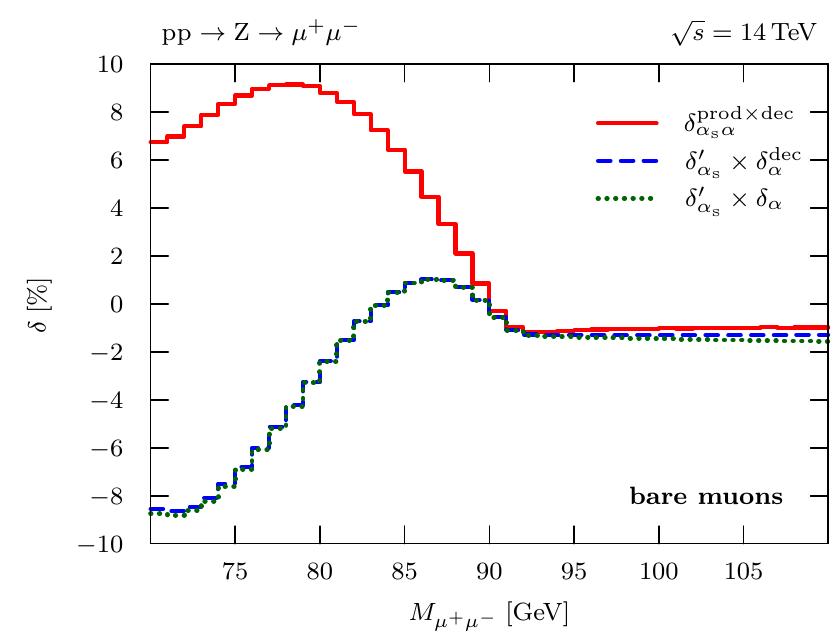}
  \includegraphics[width=.5\textwidth]{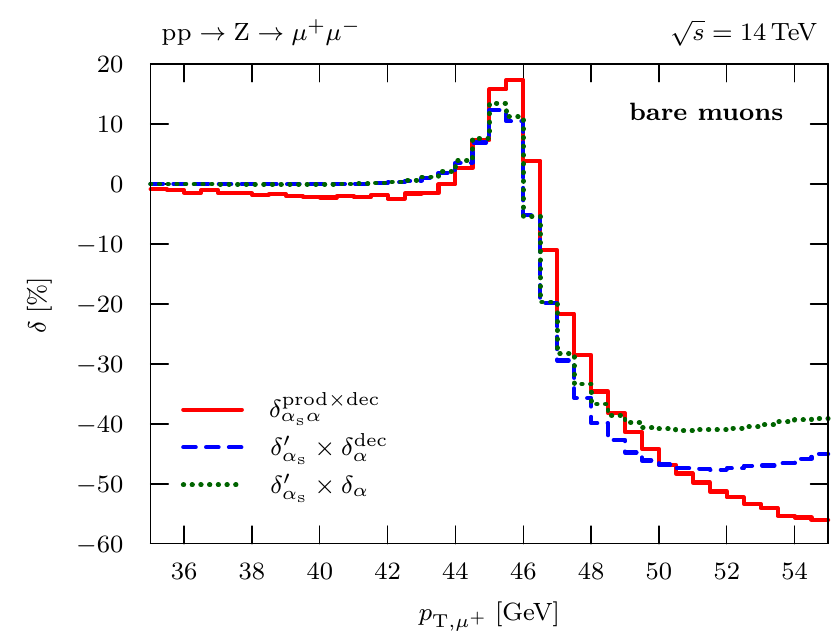}
  \caption{Relative factorizable corrections of $\order{\alphas\alpha}$ induced by initial-state QCD and final-state EW contributions to the lepton-invariant-mass distribution~(left) and a transverse-lepton-momentum
distribution~(right) for \PZ~production at the LHC.
  The naive products of the NLO correction factors $\delta_{\alphas}'$ and $\delta_\alpha$ are shown for comparison. (Taken from Ref.~\cite{Dittmaier:2015rxo}.)}
  \label{fig:distNNLO-IF-Z}
\end{figure}

Figure~\ref{fig:distNNLO-IF-Wp} shows the numerical results for the
relative \order{\alphas\alpha} initial--final factorizable corrections $\delta^{\pro\times\dec}_{\alphas\alpha}$ to the transverse-mass ($M_{\rT,\nu\Pl}$) and the
transverse-lepton-momentum ($p_{\rT,\Pl}$) distributions for \PWp\
production at the LHC.
The results for the neutral-current process are given in Fig.~\ref{fig:distNNLO-IF-Z}, which displays the results for the lepton-invariant-mass ($M_{\Pl\Pl}$) distribution and a transverse-lepton-momentum ($p_{\rT,\Pl^+}$) distribution.

In Figs.~\ref{fig:distNNLO-IF-Wp} and~\ref{fig:distNNLO-IF-Z} we
compare to the two different implementations of a naive
product of correction factors discussed after
Eq.~\eqref{eq:diff-naive}.
In both figures, we assume that final-state leptons and collinear photons
can be resolved completely (defining ``bare leptons''),
a situation that is realistic for muons, but not for
electrons, which appear in showers together with the photons
in the electromagnetic calorimeter in the detector.
For the latter case, results based on some recombination of leptons with collinear
photons are more realistic (defining ``dressed leptons'').
Such results can be found in Ref.~\cite{Dittmaier:2015rxo}.
They show the same features as the ones for bare leptons, with corrections
that are typically smaller by a factor of two.

For the $M_{\rT,\nu\Pl}$ distribution for $\PWp$ production (left plot in
Fig.~\ref{fig:distNNLO-IF-Wp}), the mixed NNLO QCD--EW corrections
are moderate and amount to approximately $\SI{-1.7}{\%}$ around the
resonance, which is about an order of magnitude smaller than the NLO
EW corrections.%
\footnote{The structure observed in the correction
$\delta^{\pro\times\dec}_{\alphas\alpha}$ around
$M_{\rT,\nu\Pl}\approx\SI{62}{\GeV}$ can be attributed to the
interplay of the kinematics of the double-real emission
corrections and the event selection. It arises close to the kinematic
boundary $M_{\rT,\nu\Pl}>\SI{50}{\GeV}$ implied by the cut
$p_{\rT,\Pl^\pm},E_{\rT}^{\text{miss}}>\SI{25}{\GeV}$ for the back-to-back kinematics
of the non-radiative process.}
Both variants of the naive product provide a good approximation to the
full result in the region around and below the Jacobian peak, which is
dominated by resonant $\PW$ production.
For larger
$M_{\rT,\nu\Pl}$, the product $\delta_{\alphas}' \delta_\alpha$
based on the full NLO EW correction factor deviates from the other curves.
Due to the well-known insensitivity of the
observable $M_{\rT,\nu\Pl}$ to ISR effects seen for the NLO corrections~\cite{Dittmaier:2014qza},
this difference signals the growing importance of effects beyond the PA.
However, the deviations amount to only few per-mille for
$M_{\rT,\nu\Pl}\lesssim\SI{90}{\GeV}$.
The overall good agreement between the
$\delta^{\pro\times\dec}_{\alphas\alpha}$ corrections and both naive
products can be attributed to well-known insensitivity of the
observable $M_{\rT,\nu\Pl}$ to ISR effects
already seen in the case of NLO corrections in Ref.~\cite{Dittmaier:2014qza}.

For the $p_{\rT,\Pl}$ distributions (right plots
in Figs.~\ref{fig:distNNLO-IF-Wp} and \ref{fig:distNNLO-IF-Z} for $\PWp$/$\PZ$ production,
respectively) we observe corrections that are small far below the
Jacobian peak, but which rise to about $15\%$ ($20\%$) on the Jacobian
peak at $p_{\rT,\Pl}\approx\MV/2$ for the case of the \PWp\ boson (\PZ\
boson) and then display a steep drop reaching almost $-50\%$ at
$p_{\rT,\Pl}=\SI{50}{\GeV}$.  This enhancement stems from the large
QCD corrections above the Jacobian peak familiar from the NLO QCD
results (see e.g.\ Fig.~8 in Ref.~\cite{Dittmaier:2014qza}) where
the recoil due to real
QCD radiation shifts events with resonant $\PW$/$\PZ$ bosons above the
Jacobian peak.
The naive product ansatz fails to provide a good description of the full
result $\delta^{\pro\times\dec}_{\alphas\alpha}$ and deviates by
5--10\% at the Jacobian peak, where the PA is expected to be the most
accurate.  This can be attributed to the strong influence of the
recoil induced by ISR on the transverse momentum,
which implies a larger effect of the double-real emission corrections
on this distribution that are not captured correctly by the naive
products.  The two versions of the naive products display larger
deviations than in the $M_{\rT,\nu\Pl}$ distribution discussed above,
which signals a larger impact of the missing \order{\alphas\alpha}
initial--initial corrections.  However, these deviations should be
interpreted with care,
since a fixed-order
prediction is not sufficient to describe this distribution around the
peak region $p_{\rT,\Pl}\approx\MV/2$, which corresponds to the
kinematic onset for $V+\text{jet}$ production and is known to
require QCD resummation for a proper description.

In case of the $M_{\Pl\Pl}$
distribution for $\PZ$ production (left plot in
Fig.~\ref{fig:distNNLO-IF-Z}),  corrections up to $10\%$  are
observed below the resonance for the case of bare muons. This is consistent with the large EW
corrections at NLO in this region, which arise from photonic final-state
radiation (FSR) that shifts the reconstructed value of the invariant
lepton-pair mass away from the resonance to lower values.
The naive product approximates the
full initial--final corrections
$\delta^{\pro\times\dec}_{\alphas\alpha}$ reasonably well at the resonance
itself ($M_{\Pl\Pl}=\MZ$) and above, but completely fails already a
little below the resonance where the naive products do not even
reproduce the sign of the full
$\delta^{\pro\times\dec}_{\alphas\alpha}$ correction.
This deviation occurs although the invariant-mass distribution is
widely unaffected by ISR effects.
The fact that we obtain almost identical corrections
from the two versions of the product $\delta_{\alphas}'\delta_\alpha^{\dec}$
and $\delta_{\alphas}'\delta_\alpha$ demonstrates the insensitivity of this
observable to photonic ISR.
The origin of the failure of the naive product ansatz, which is discussed
in Ref.~\cite{Dittmaier:2015rxo} in detail, can be understood as follows.
The large EW corrections below the resonance
arise due to the redistribution of events near the $\PZ$
pole to lower lepton invariant masses by photonic FSR, so that
it would be more appropriate to replace
the QCD correction factor $\delta_{\alphas}'$ in the naive product by its value at the resonance
$\delta_{\alphas}'(M_{\Pl\Pl}=\MZ)\approx 6.5\%$, which corresponds to
the location of the events that are responsible for the bulk of the
large EW corrections below the resonance.
In contrast, the naive product ansatz simply multiplies the
corrections locally on a bin-by-bin basis.
The observed mismatch is dramatically enhanced by the fact that the
QCD correction $\delta_{\alphas}'$ exhibits a sign change at
$M_{\Pl\Pl}\approx\SI{83}{\GeV}$.

Contrary to the lepton-invariant-mass distribution, the
transverse-mass distribution is dominated by events with resonant
W~bosons even in the range below the Jacobian peak,
$M_{\rT,\nu\Pl}\lesssim\MW$, so that it is less sensitive to the
redistribution of events to lower $M_{\rT,\nu\Pl}$.
This explains why the naive product can provide a good approximation
of the full initial--final NNLO corrections.  It should be emphasized,
however, that even in the case of the $M_{\rT,\nu\Pl}$ distribution
any event selection criteria that deplete events with resonant W~bosons
below the Jacobian peak will result in increased sensitivity to the effects of
FSR and can potentially lead to a failure of a naive
product ansatz.

In conclusion, simple approximations in terms of products of
correction factors have to be used with care and require a careful
case-by-case investigation of their validity.

\subsection{Approximating \order{\alphas\alpha} corrections by
leading logarithmic final-state radiation}

As is evident from Figs.~\ref{fig:distNNLO-IF-Wp} and~\ref{fig:distNNLO-IF-Z}, a naive
product of QCD and EW correction factors~\eqref{eq:def:nnlo:naive:fact} is not
adequate to approximate the NNLO QCD--EW corrections for all observables.
A promising approach to a factorized approximation for the dominant initial--final corrections
can be obtained by combining the full NLO QCD corrections to vector-boson production
with the leading-logarithmic (LL) approximation for FSR.
The benefit in this approximation lies in the fact that the interplay of the recoil
effects from jet and photon emission is properly taken into account.
On the other hand, the LL approximation neglects certain (non-universal)
finite contributions, which are, however, suppressed with respect to the dominating radiation effects.

In the following we compare two standard approaches to include FSR off leptons in
LL approximation: the structure-function and the parton-shower approaches.
In the former, generated events are dressed with FSR effects by convoluting the
differential cross section by a structure function describing the energy loss
of the leptons by collinear photon emission.
Since the structure-function approach works with strictly collinear photon emission,
by construction the impact of LL FSR is zero for dressed leptons,
where the mass-singluar logarithm of the lepton cancels by virtue of the
KLN theorem because of the inclusive treatment of the collinear lepton--photon system.
In contrast, photons generated through parton-shower approaches to photon radiation
(see e.g.\ Refs.~\cite{Placzek:2003zg,CarloniCalame:2003ux,CarloniCalame:2005vc})
also receive momentum components transverse to the original lepton momentum,
following the differential factorization formula, so that
the method is also applicable to the case of collinear-safe observables,
i.e.\ to the dressed-lepton case. For this purpose, we have
implemented the combination of the exact NLO QCD prediction for
vector-boson production with the simulation of FSR
using PHOTOS~\cite{Golonka:2005pn}. Since we are interested in
comparing  to the
\order{\alphas\alpha} corrections in our setup, we only generate a
single photon emission using PHOTOS and use the same input-parameter scheme for
$\alpha$ (see Ref.~\cite{Dittmaier:2015rxo} for details).

\begin{figure}[t]
  \includegraphics[width=.5\textwidth]{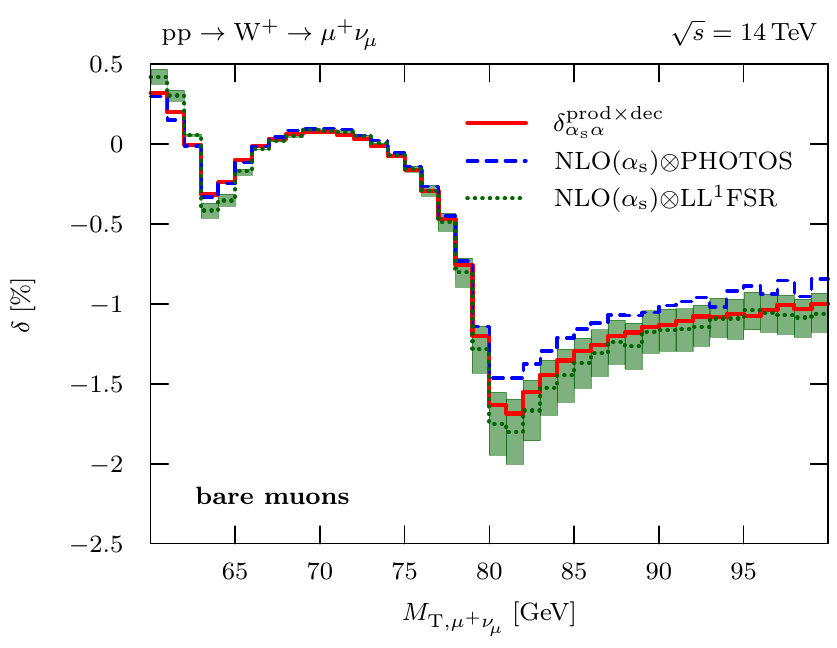}
  \includegraphics[width=.5\textwidth]{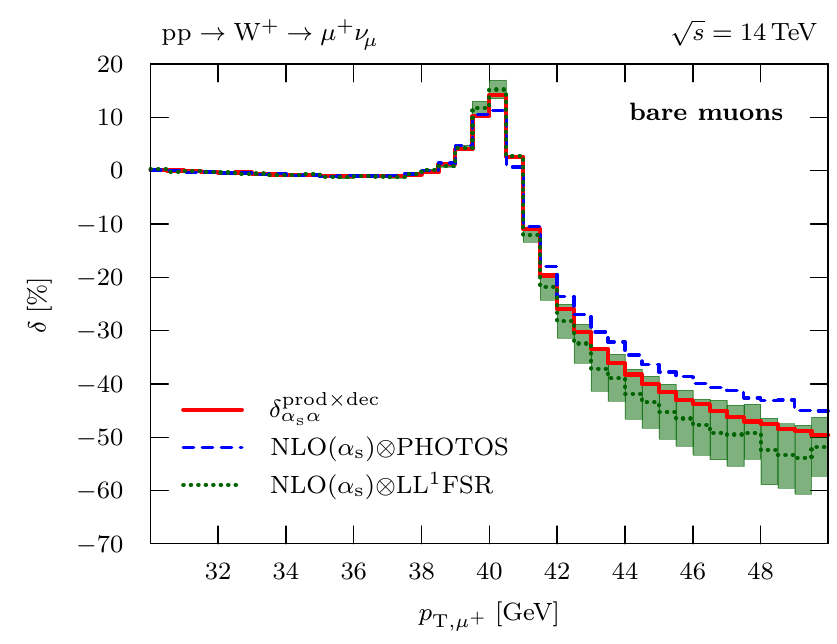}
  \caption{Comparison of the approximation obtained
    from PHOTOS and from the structure-function approach ($\mathrm{LL}^1\mathrm{FSR}$)
    for the relative $\order{\alphas\alpha}$ initial-state
    QCD and final-state EW corrections to our best prediction
    $\delta^{\mathrm{prod\times dec}}_{\alphas\alpha}$ for the case of the
    transverse-mass~(left) and transverse-lepton-momentum~(right)
    distributions for \PWp~production at the LHC. (Taken from Ref.~\cite{Dittmaier:2015rxo}.)}
  \label{fig:distNNLO-IF-FSR-Wp}
%
\vspace{2.5em}
  \includegraphics[width=.5\textwidth]{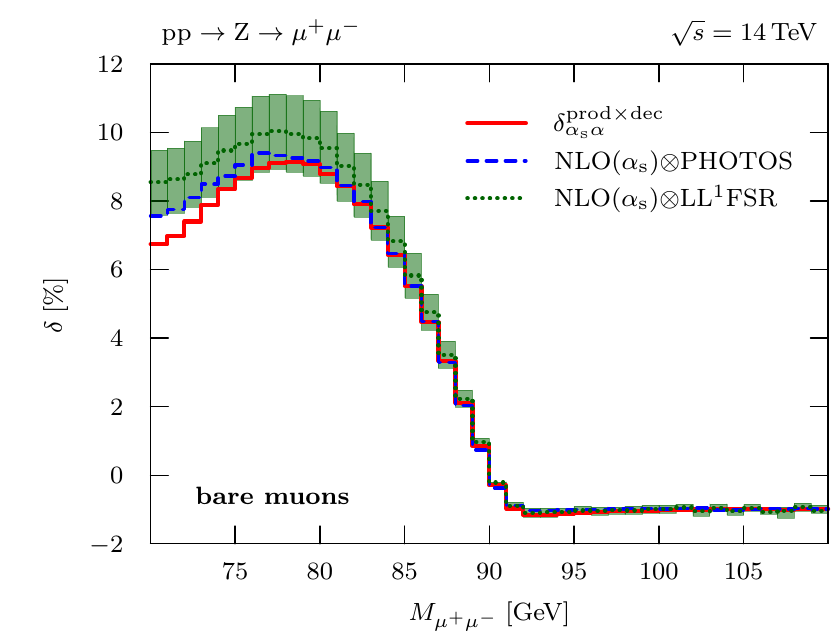}
  \includegraphics[width=.5\textwidth]{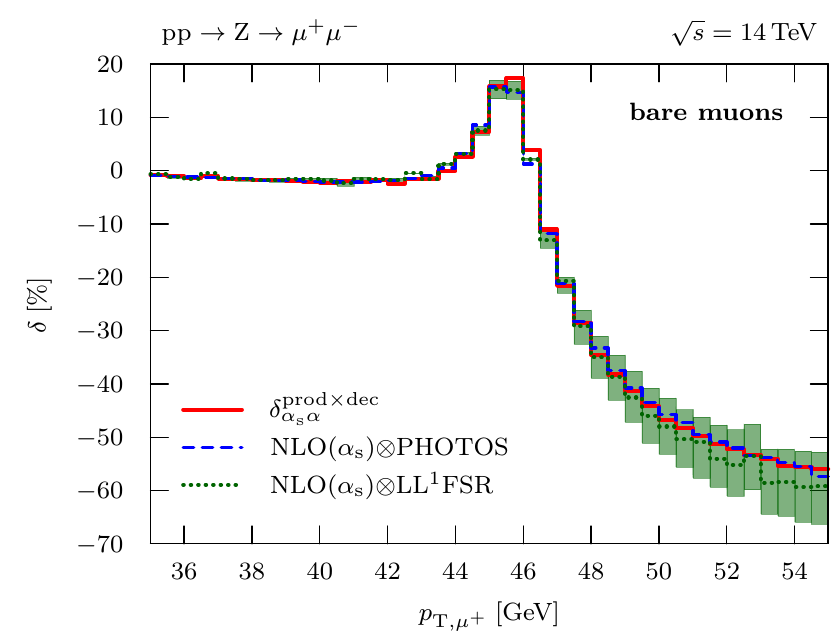}
  \caption{Comparison of the approximation obtained from PHOTOS
    and from the structure-function approach ($\mathrm{LL}^1\mathrm{FSR}$) for the
    relative $\order{\alphas\alpha}$ initial-state QCD and final-state
    EW corrections to our best prediction
    $\delta^{\mathrm{prod\times dec}}_{\alphas\alpha}$ for the case of the
    lepton-invariant-mass distribution~(left) and a
    transverse-lepton-momentum distribution~(right) for \PZ~production at
    the LHC. (Taken from Ref.~\cite{Dittmaier:2015rxo}.)}
  \label{fig:distNNLO-IF-FSR-Z}
\end{figure}

In Figs.~\ref{fig:distNNLO-IF-FSR-Wp} and \ref{fig:distNNLO-IF-FSR-Z}
we compare our best prediction~\eqref{eq:def:nnlo:if} for the
factorizable initial--final \order{\alphas\alpha} corrections to the
combination of NLO QCD corrections with the approximate FSR
obtained from the structure-function approach and PHOTOS for the case of $\PWp$ production and $\PZ$
production, respectively.
The corresponding results for dressed leptons can be found in Ref.~\cite{Dittmaier:2015rxo}.
The combination of the NLO QCD corrections and
approximate FSR leads to a clear improvement compared to the
naive product approximations investigated in the previous section.
This is particularly apparent in the
neutral-current process where the $M_{\Pl\Pl}$ distribution is correctly modelled
by both LL FSR approximations, whereas the naive products shown in
Figs.~\ref{fig:distNNLO-IF-Z} completely failed to
describe this distribution.
In the $M_{\rT,\nu\Pl}$ spectrum of the charged-current
process in Fig.~\ref{fig:distNNLO-IF-FSR-Wp} one also finds good
agreement of the different results below the Jacobian peak and an
improvement over the naive product approximations in
Fig.~\ref{fig:distNNLO-IF-Wp}.
The description of the $p_{\rT,\Pl}$ distributions is also improved compared to the naive product approximations, but some differences remain in the charged-current process.

In spite of the good agreement of the two versions of incorporating FSR effects,
the intrinsic uncertainty of the LL approximations should be kept in mind.
For the structure-function approach, this uncertainty is illustrated by the band width
resulting from the uncertainty in the QED scale $Q$, which is not determined at the LL level
and varied within the range $M_V/2<Q<2M_V$ for $V=\PW,\PZ$.
We remark that the multi-photon corrections obtained by employing the
structure function with effects beyond \order{\alpha}
lie well within the aforementioned scale bands, which shows that a
proper matching to the full NLO EW calculation is needed to
remove the dominant uncertainty of the LL approximation and to
predict the higher-order effects reliably.
For PHOTOS the intrinsic uncertainty is not shown and not easy to
quantify.
The good quality of the PHOTOS approximation results from the fact that the finite terms in the photon emission probability are specifically adapted to $\PW/\PZ$-boson decays.
The level of agreement with our ``full prediction'', thus, cannot be taken over to other processes.

\subsection{Impact on the W-boson mass extraction}
\label{sec:mw}

In order to estimate the effect of the \order{\alphas \alpha}
corrections on the extraction of the $\PW$-boson mass at the LHC we
have performed a $\chi^2$ fit of the $M_{\rT,\nu \ell}$ distribution.
We treat the $M_{\rT,\nu \ell}$ spectra calculated in various
theoretical approximations for a reference mass $\MW^{\OS}= 80.385~\GeV$
as ``pseudo-data'' that we fit with
 ``templates'' calculated
using the LO predictions $\sigma^0$ (with NLO PDFs)
for different values of $\MW^{\OS}$.
Specifically, we have generated results for $27$ transverse-mass bins in the
interval $M_{\rT,\nu \ell}= [64,\,91]~\GeV$ in steps of $1~\GeV$, varying the $\PW$-boson mass in the interval $\MW=[80.085,\,80.785]~\GeV$ with steps of $\Delta \MW=10~ \MeV$ (steps of $\Delta
\MW=5~\MeV$ in the interval $\MW=[80.285,\,80.485]~\GeV$).
Using a linear interpolation between neighbouring $\MW$ values, we obtain the integrated cross sections in the $i$-th $M_{\rT,\nu \ell}$ bin, $\sigma_i^0(\MW)$, as a continuous function of $\MW$.
The best-fit value $\MW^{\mathrm{fit}, \mathrm{th}}$
quantifying the impact of a higher-order correction
in the theoretical cross section $\sigma^{\mathrm{th}}$ is
then obtained from the minimum of the function
\begin{equation}
\label{eq:chi2}
  \chi^2(\MW^{\mathrm{fit},\mathrm{th}})=\sum_{i}
  \frac{\left[\sigma_i^{\mathrm{th}}(\MW^{\OS})
      -\sigma_i^0(\MW^{\mathrm{fit},\mathrm{th}})\right]^2}{
    2 \Delta\sigma_i^2}\,,
\end{equation}
where the sum over $i$ runs over the transverse-mass bins.
Here $\sigma_i^{\mathrm{th}}$ and $\sigma_i^0$ are the
integrated cross sections in the $i$-th bin, uniformly rescaled so
that the sum over all $27$ bins is identical for all considered cross
sections.  We assume a statistical error of the pseudo-data and take
$\Delta\sigma_i^2\propto \sigma_i^{\mathrm{th}}$.  We have also
performed a two-parameter fit where the normalization of the templates
is fitted simultaneously, leading to identical results.  Similarly,
allowing the $\PW$-boson width in the templates to float and fitting
$\MW$ and $\GW$ simultaneously does not
significantly affect our results on the mass shift.

In the experimental measurements of the transverse-mass distribution,
the Jacobian peak is washed out due to the finite energy and momentum
resolution of the detectors. In our simple estimate of the impact of
higher-order corrections on the extracted value of the $\PW$-boson
mass, we do not attempt to model such effects.  We expect the detector
effects to affect the different theory predictions in a similar way
and to cancel to a large extent in our estimated mass shift, which is
obtained from a difference of mass values extracted from pseudo-data
calculated using different theory predictions.  This assumption is
supported by the fact that our estimate of the effect of the NLO EW
corrections is similar to the one obtained in
Ref.~\cite{CarloniCalame:2003ux} using a Gaussian smearing of the
four-momenta to simulate detector effects.

\begin{table}
\centering
\begin{tabular}{@{} l r@{\enspace}l r@{\enspace}l @{}}
  \toprule
  & \multicolumn{2}{c}{bare muons} & \multicolumn{2}{c}{dressed leptons} \\
  \cmidrule(r){2-3} \cmidrule(l){4-5}
  & $M^\mathrm{fit}_\mathrm{W}\;[\mathrm{GeV}]$ & $\qquad\Delta M_\mathrm{W}$
  & $M^\mathrm{fit}_\mathrm{W}\;[\mathrm{GeV}]$ & $\qquad\Delta M_\mathrm{W}$ \\
  \midrule
  LO          & $80.385$ & \multirow{2}*{$\bigg\}\;-90~\mathrm{MeV}$}
              & $80.385$ & \multirow{2}*{$\bigg\}\;-40~\mathrm{MeV}$} \\
         $\NLO_\rew$ & $80.295$ &
              & $80.345$ &                   \\
  \midrule
   $\NLO_{\rs\oplus\rew}$
        & $80.374$ & \multirow{2}*{$\biggr\}\;-14~\mathrm{MeV}$}
              & $80.417$ & \multirow{2}*{$\biggr\}\;-4~\mathrm{MeV}$} \\
  NNLO        & $80.360$ &
              & $80.413$ &                   \\
  \bottomrule
\end{tabular}
\caption{Values of the $\PW$-boson mass in GeV obtained from the $\chi^2$ fit
of the  $M_{\rT,\nu \ell}$ distribution in  different theoretical
approximations  to LO templates and the resulting mass shifts. (Taken from Ref.~\cite{Dittmaier:2015rxo}.)}
  \label{tab:mass_shift}
\end{table}

The  fit results for several NLO approximations and our best NNLO prediction~\eqref{eq:def:nnlo:if} are given in Table~\ref{tab:mass_shift}.
To validate our procedure we estimate the mass shift due to the NLO EW
corrections by using the prediction
$\sigma^{\NLO_\rew}=\sigma^0+\Delta
\sigma^{\NLO_\rew}$
as the pseudo-data $\sigma^{\mathrm{th}}$
in~\eqref{eq:chi2}. The corresponding $\chi^2$ distributions
as a function of the mass shift $\Delta\MW^{\NLO_\rew}$
can be found in Ref.~\cite{Dittmaier:2015rxo}.
From the minima of the distributions one finds a mass shift of
$\Delta \MW^{\NLO_\rew}\approx -90~\MeV$ for bare muons and $\Delta
\MW^{\NLO_\rew}\approx -40~\MeV$ for dressed muons.  These values are
comparable to previous results reported in
Ref.~\cite{CarloniCalame:2003ux}.\footnote{
  In Ref.~\cite{CarloniCalame:2003ux} the values $\Delta\MW=110~\MeV$
  ($20~\MeV$) are obtained for the bare-muon (dressed-lepton)
  case. These values are obtained using the
  $\order{\alpha}$-truncation of a LL shower and for
  lepton-identification criteria appropriate for the Tevatron taken
  from Ref.~\cite{Baur:1998kt}, so they cannot be compared directly to
  our results. In particular, in the dressed-lepton case, a looser
  recombination criterion $R_{\Pl^\pm\gamma}< 0.2$ is applied, which is
  consistent with a smaller impact of the EW corrections.  Note that
  the role of pseudo-data and templates is reversed in
  Ref.~\cite{CarloniCalame:2003ux} so that the mass shift has the
  opposite sign.}
Alternatively, the effect of the EW corrections can
be estimated by comparing the value of $\MW$ obtained from a fit to
the naive product of EW and QCD
corrections~\eqref{eq:def:nnlo:naive:fact} to the result of a fit to
the NLO QCD cross section. The results are consistent with the shift
estimated from the NLO EW corrections alone.

We have also estimated the effect of multi-photon radiation on the
$\MW$ measurement in the bare-muon case using the
structure-function approach.
We obtain a mass shift $\Delta\MW^{\mathrm{FSR}}\approx 9~\MeV$
relative to the result of the fit to the NLO EW prediction, which
is in qualitative agreement with the result of
Ref.~\cite{CarloniCalame:2003ux}.

To estimate the impact of the  initial--final \order{\alphas\alpha}
corrections  we consider the mass shift relative to the full NLO result,
\begin{equation}
  \label{eq:dmwnnlo}
  \Delta\MW^{\NNLO}=\MW^{\mathrm{fit},\NNLO_{\rs\otimes\rew}^{\pro\times\dec}}
  -\MW^{\mathrm{fit},\NLO_{\rs\oplus\rew}},
\end{equation}
where $\MW^{\mathrm{fit},\NNLO_{\rs\otimes\rew}^{\pro\times\dec}}$ is the
result of using our best
prediction~\eqref{eq:def:nnlo:if} to generate the pseudo-data, while
the sum of the NLO QCD and EW corrections is used for
$\Delta\MW^{\mathrm{fit},\NLO_{\rs\oplus\rew}}$. The resulting $\chi^2$
distributions for the mass shift can again be found in
Ref.~\cite{Dittmaier:2015rxo}.
In the bare-muon case, we obtain
a mass shift due to \order{\alphas\alpha} corrections of $\Delta
\MW^{\mathrm{NNLO}}\approx -14~\MeV$, while for the dressed-lepton case we
get $\Delta \MW^{\mathrm{NNLO}}\approx -4~\MeV$.

Identical shifts result from replacing the NNLO prediction by the
naive product~\eqref{eq:def:nnlo:naive:fact}, which is expected from
the good agreement for the $M_{\rT,\nu \ell}$-spectrum in
Fig.~\ref{fig:distNNLO-IF-Wp}.  Using instead the LL
approximation of the FSR obtained using
PHOTOS to compute the \order{\alphas\alpha} corrections, we obtain a
mass shift of $\Delta\MW^{\NNLO}=-11~\MeV$ ($-4~\MeV$) for the
bare-muon (dressed-lepton) case.  The effect of the
\order{\alphas\alpha} corrections on the mass measurement is therefore
of a similar or larger magnitude than the effect of multi-photon
radiation.
  We emphasize that
the result $\Delta\MW^{\NNLO}\approx -14~\MeV$ is a simple estimate of the
 impact of the full \order{\alphas\alpha} corrections
on the $\MW$ measurement.  The order of magnitude shows that these
corrections must
be taken into account properly in order to reach the $10~\MeV$ accuracy
goal of the LHC experiments.

\section{Conclusions}

The Drell--Yan-like $\PW$- and $\PZ$-boson production processes are among the
most precise probes of the Standard Model and do
not only serve as key benchmark or ``standard candle'' processes,
but further allow for precision measurements of the $\PW$-boson mass
and the effective weak mixing angle.
In view of the envisioned accuracy of these measurements, a further improvement of the theory prediction is mandatory.
To this end, the mixed QCD--electroweak corrections of
\order{\alphas\alpha} represent the largest component of fixed-order
radiative corrections after the well established NNLO QCD and NLO electroweak
corrections.

In this article, we have reviewed the major results of
our two recent papers~\cite{Dittmaier:2014qza,Dittmaier:2015rxo}, where
we have established a framework for evaluating the \order{\alphas\alpha} corrections
to Drell--Yan processes in the resonance region using a pole approximation and
presented the calculation of the non-factorizable and most important factorizable
corrections.
The non-factorizable corrections and the factorizable corrections solely associated with the W/Z~decay subprocesses, which were computed in
Ref.~\cite{Dittmaier:2014qza} and \cite{Dittmaier:2015rxo}, respectively,
turned out to be phenomenologically negligible.
Moreover, an analysis of the NLO corrections in pole approximation suggests that
the factorizable corrections corresponding to the production subprocess, which are
yet unknown, will have a minor impact on the observables relevant for the
W-boson mass measurement.
The dominant factorizable corrections of $\order{\alphas\alpha}$ are the ones resulting from
the combination of sizable QCD corrections to the production with large EW corrections to the decay subprocesses,
explicitly calculated in Ref.~\cite{Dittmaier:2015rxo}.
Here we have summarized the numerical results for so-called bare leptons
and the most important observables for the $\PW$-boson mass measurement:
the transverse-mass and lepton-transverse-momentum distributions for \PW\ production.
The results for the neutral-current process comprise the invariant-mass and
the lepton-transverse-momentum distributions.

We have shown that naive products fail to capture the factorizable
initial--final corrections in distributions such as in
the transverse momentum of the lepton, which are sensitive to QCD
initial-state radiation and therefore require  a correct treatment of the double-real-emission
part of the NNLO corrections. Naive products also fail to capture observables that are strongly
affected by a redistribution of events due to final-state real-emission corrections,
such as the invariant-mass distribution of the neutral-current process.
On the other hand, if an observable is less affected by such a redistribution of events or
is only affected by it in the vicinity of the resonance, such as the transverse-mass
distribution of the charged-current process, the naive products are able to reproduce
the factorizable initial--final corrections to a large extent.
Moreover, we have investigated to which extent the factorizable initial--final corrections
calculated in this paper can be approximated by a combination of the NLO QCD corrections
and a collinear approximation of  real-photon emission through a  QED structure-function approach
or a  QED parton shower such as PHOTOS.
For the invariant-mass distribution in $\PZ$-boson production we observe a significant
improvement in the agreement compared to the naive product ansatz, since
both PHOTOS and the QED structure function model the redistribution of events due to
final-state radiation, which is responsible for the bulk of the corrections in this observable.
Our results can furthermore be used to validate Monte Carlo event generators where
$\order{\alphas\alpha}$ corrections are approximated by a combination of NLO matrix
elements and parton showers.

Finally,
we have illustrated the phenomenological impact of the $\order{\alphas\alpha}$ corrections
by estimating the mass shift induced by the factorizable initial--final corrections as
$\approx-14~\MeV$  for the case of bare muons  and $\approx-4~\MeV$ for dressed leptons.
These corrections therefore have to be properly taken into account in the $\PW$-boson mass
measurements at the LHC, which aim at a precision of about $10~\MeV$.
It will be interesting to investigate the impact of the $\order{\alphas\alpha}$
corrections on the measurement of the effective weak mixing angle as well in the future.

\section*{Acknowledgement}
This project is supported by the German Research Foundation (DFG) via grant DI 784/2-1 and the German Federal Ministry for Education and Research (BMBF).
Moreover, A.H.\ is supported via the ERC Advanced Grant MC@NNLO (340983).
C.S.\ is supported by the Heisenberg Programme of the Deutsche Forschungsgemeinschaft.

\end{document}